\documentclass[runningheads]{cl2emult}
\input epsf

\def\msol{M_\odot}
\def\mj{M_J}
\def\lsol{L_\odot}

\def\mvol{{\rm M}_\odot{\rm pc}^{-3}}
\def\msurf{{\rm M}_\odot{\rm pc}^{-2}}

\def\te{T_{\rm eff}}

\def\simgr{\,\hbox{\hbox{$ > $}\kern -0.8em \lower 1.0ex\hbox{$\sim$}}\,}
\def\simle{\,\hbox{\hbox{$ < $}\kern -0.8em \lower 1.0ex\hbox{$\sim$}}\,}
\def\etal{{\it et al.}}

\def\aj{AJ}                  % Astronomical Journal
\def\araa{ARA\&A}             % Annual Review of Astron and Astrophys
\def\apj{ApJ}                 % Astrophysical Journal
\def\apjl{ApJ}                % Astrophysical Journal, Letters
\def\apjs{ApJS}               % Astrophysical Journal, Supplement
\def\aap{A\&A}                % Astronomy and Astrophysics
\begin{document}

\title*{Evolution of low-mass stars and substellar objects.
 Contribution to the Galactic mass budget}
\toctitle{Evolution of low-mass stars and substellar objects.
\protect\newline
 Contribution to the Galactic mass budget}
\titlerunning{Evolution of low-mass stars and substellar objects}

\author{G. Chabrier\inst{1}
\and I. Baraffe\inst{1}
\and F. Allard\inst{1}
\and P.H. Hauschildt\inst{2}}

\authorrunning{G. Chabrier et al.}

\institute{Centre de Recherche Astrophysique de Lyon (UMR CNRS 5574),\\
 Ecole Normale Sup\'erieure de Lyon, 69364 Lyon
Cedex 07, France
\and  Dpt. of Physics \& Astronomy, University of Georgia, Athens, GA 30602, USA}

\maketitle

\begin{abstract}
We briefly summarize our present knowledge of the theory of low-mass stars and substellar objects and their contribution to
the Galactic population.
\end{abstract}

\section{ Introduction}
 
The search for substellar objects (SSO) has bloomed over the past few years
with the unambiguous identification of several free floating brown dwarfs (BDs)
and with the discovery of numerous planets orbiting stars outside the solar system.
Substellar objects therefore exist and ongoing and future observational projects
are likely
to reveal dozens more of such objects.
Two groups, the Lyon group and the Tucson group, have aspired to derive a complete theory of
the evolution and the spectral signature of low-mass, dense
objects, from Sun-like to Saturn-like masses, covering 3 orders of magnitude in
mass and 9 in luminosity, bridging the gap between stars and gaseous planets
(see e.g. Burrows \etal, 1997 and Baraffe \etal, 1998 and references therein).
These two groups have incorporated the best possible physics aimed at describing the
mechanical and thermal properties of these objects - equation of state, synthetic spectra,
non-grey atmosphere models. Both groups have "met in the middle" by successfully identifying the main spectral properties
of the benchmark BD Gl229B, providing a determination of its mass and age (Marley \etal, 1996; Allard \etal, 1996). We refer to the afore-mentioned papers and the references therein for a complete description of the
physics entering these models. Only a brief outline of these characteristics is
given below.

\section{The physics of sub-stellar objects}

\subsection{Interior}

Central conditions for SSOs are typically $T_c\simle 10^5$ K and
$\rho_c\sim 10^2$-$10^3$ g cm$^{-3}$, characterizing a strongly coupled electron-ion plasma. Moreover, in the envelope, the electron binding energy
is of the order of the Fermi energy $Ze^2/a_0\sim E_F$ so that {\it pressure}-dissociation
and ionization take place along the interior. Very recently laser-driven shock wave
experiments at Livermore have achieved pressures on liquid D$_2$ of up to 5 Mbar at high temperature
(Collins \etal, 1998), exploring
for the first time the regime of pressure-dissociation and ionization, therefore probing
the equation of state (EOS) under conditions characteristic of SSO interiors. As shown in Figure 3 of Collins \etal,
the experimental hugoniot revealed the excellent behaviour of the EOS developed by Saumon and
Chabrier (Saumon \etal, 1995 and references therein). In particular the pronounced compressibility observed in the
dissociation domain ($\rho/\rho_0=5.88)$ agrees remarkably well with the theoretically
predicted value. These experiments open a new window in physics and astrophysics by
constraining the physics of the interior of SSOs in laboratory experiments.

\subsection{Atmosphere}

In the atmosphere, the hydrostatic equilibrium condition $d\tau = \bar \kappa dP/g$, where $g=Gm/R^2\approx 10^3$-$10^5$
cm s$^{-2}$ is the characteristic surface gravity of SSOs, yields $P_{ph}\sim g/\bar \kappa
\approx $0.1-10 bar and $\rho_{ph}\approx$10$^{-6}$-10$^{-4}$ g cm$^{-3}$ at the photosphere. Collisional effects are important under such conditions and introduce sources of absorption
like e.g. the collision-induced absorption (CIA) between H$_2$-H$_2$ or H$_2$-He below $\sim 5000$ K (see e.g. Borysow \etal, 1985).
In the effective temperature range characteristic of low-mass stars (LMS) and
SSOs ($\te \simle 5000$ K), numerous molecules
form, in particular metal oxydes and hydrides (TiO,VO,FeH,CaH), the major absorbers in the optical,
and CIA H$_2$, H$_2$O, CO which dominate in the infrared (see Allard \etal, 1997 for a review).
The situation becomes even more complicated for SSOs,
due to the changes in molecular chemistry
which occur across the atmospheric temperature range from 2000 K (the
coolest stars) to 170 K (jovian conditions).  At 2000 K, most  of
the carbon is locked into carbon monoxide CO, while the oxygen is found in water vapor H$_2$O, dominantly, and in
titanium TiO and vanadium VO monoxides.
Below $\sim 1800$ K, the dominant equilibrium form of carbon is no longer CO but CH$_4$ (Fegley \& Lodders, 1996). As confirmed by the observation of Gl229B (Oppenheimer \etal, 1995), methane features begin to appear in the infrared while
titanium dioxide and silicate clouds form at the expense of TiO, modifying
profoundly the opacity of the atmosphere.
For jovian-like atmospheres, the dominant equilibrium form of
nitrogen is NH$_3$ ($\te \simle 600$ K) and below $\te \sim 200$ K water
condenses to clouds at or above the photosphere.
As shown by Tsuji \etal (1996) and
Jones \& Tsuji (1997), there is evidence for condensation of metals and silicates into grains (e.g. TiO into CaTiO$_3$, Mg, Si into
MgSiO$_3$) for $\te \simle 3000$ K,
i.e. at the bottom of the main sequence.

A proper inclusion of this complex atmospheric chemistry, and of a proper calculation of unknown cross-sections, represent the main challenge
for a correct description of the spectral signature and
the evolution of substellar objects.
This grain formation process has been included in the calculations of the Tucson group only
in term of condensation. The species are precipitated according to their condensation curves.
If a species has condensed, it is left at its saturated vapor pressure (Burrows \etal, 1997).
There is no inclusion of the radiative transfer effect of the grains. These
models can thus be considered as grainfree opacity models. Recently, the Lyon group has
extended similar calculations by implicitely including all condensed species into the radiative
transfer equations (see Allard \& Hauschildt, these proceedings). The atmosphere profiles were matched with the interior
profiles at an optical depth deep enough to lie on the internal adiabat (see Chabrier \& Baraffe, 1997).

\section{Color-magnitude diagram}

The present evolutionary calculations have been conducted with three different atmosphere models: (i) for hot objects, i.e.
massive or young enough, to
preclude the formation of grains ($\te \simgr 2800$ K), we have used the most recent grainless
NGEN atmosphere models (Hauschildt \etal, 1999); (ii) for objects in the range 3000-1000 K, i.e. from
the bottom of the MS down to Gliese229B-like objects, we have used complete atmosphere models
wich include the grain opacity sources in the transfer equations,
the so-called DUSTY models (Allard \& Hauschildt, these proceedings);
(iii) for objects below 3000 K down to jovian temperatures, we have also considered
cases where the condensates settle rapidly below the photosphere and - although modifying
the atmosphere EOS - do not participate to the opacity, the so-called COND models.
This is similar to the Burrows \etal (1997) calculations and is
motivated by the relative absence of grain features in the atmosphere of objects below $\te \sim 1000$ K,
i.e. Gliese229B-like and cooler objects.
We found the effect
of grain opacity (DUSTY) to affect only moderately the H-burning minimum mass. Models with grainless atmosphere yield
$m=0.072\,\msol$, $L=5\times 10^{-5}\,\lsol$ and $\te=1700$ K at the H-burning limit, whereas models with grain
opacity give $m\approx\,0.07\msol$, $L\approx 4\times 10^{-5}\,\lsol$ and $\te\approx 1600$ K, for solar composition.

\begin{figure}
%\centering
%\includegraphics[width=.4\textwidth]{CMD-KJK.eps}
\begin{center}
\epsfxsize=85mm
\epsfysize=85mm
\epsfbox{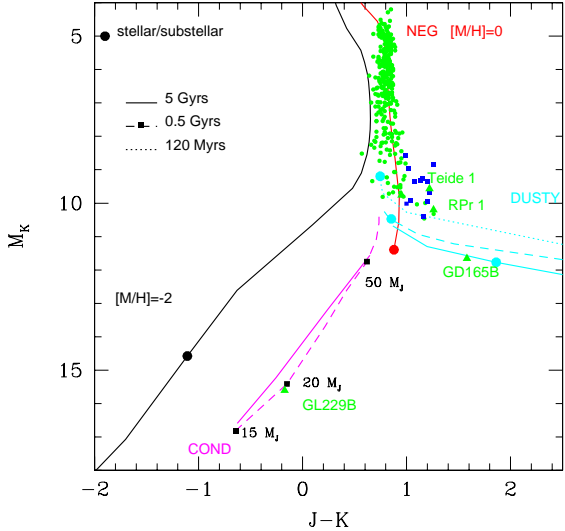}
\end{center}
%\sidecaption\mpicplace{width cm}{height cm}
\caption[]{M$_K$ vs (J-K) diagram for 3 isochrones, namely 5$\times 10^9$ yrs,
5$\times 10^8$ yrs and 1.2$\times 10^8$ yrs (the age of the Pleiades). The small circles and squares correspond to the
observation of main sequence stars (Leggett, 1996) and of Pleiades objects (Bouvier \etal, 1998), respectively. Some identified BDs are also indicated}
\label{eps1}
\end{figure}

Figure 1 displays a K vs J-K color-magnitude diagram (CMD).
In terms of colors there is a competing effect between grain and molecular opacity sources for objects at the bottom and just below the MS. 
As already identified for e.g. Gl229B (Allard et al., 1996; Marley et al., 1996) CH$_4$ and
CIA of H$_2$ lead to bluer infrared colors for
SSOs, as illustrated by the NEG and COND models, whereas grain opacity
results in a severe redening. Indeed, the present DUSTY models reproduce the spectra of the DENIS objects (Tinney \etal, 1998) and of the long puzzling object GD165B, which lies at the very edge of the H-burning limit (Kirkpatrick \etal, 1999). As $\te$ decreases, grains settle below the photosphere and the DUSTY track will merge with the COND tracks. Observations of objects in the region between these two extreme cases will be the next
confirmation of the present theory.
In spite of this strong
absorption in the IR, however,
BDs around 1500 K radiate nearly 90\% (99\% with dust) of their
energy at wavelengths longward of $1 \, \mu$m  and infrared
colors are still preferred to optical colors (at least for solar metal abundance), with $M,L,K$ as the favored bands, and M$_M\sim$M$_L\sim11$, M$_K\sim 12$ at the H-burning limit, at 5 Gyr.
Direct observation of the characteristic fluxes of SSO's are now within
reach with several observational projects, like SOFIA, SIRTF, ISAAC
(Burrows \etal, 1997; Allard 
\& Hauschildt, these proceedings).

Several BD surveys are presently conducted in young clusters and it is important to develop
accurate (non-grey) models for pre-MS stars and young BD's. The more massive of such
objects will be hot enough so that grain formation does not occur, but for $t\simle 5\times 10^7$ yrs, proper evolutionary
calculations must consider two effects : (i) first of all the influence of the initial conditions must be carefully examined because they will significantly affect the mass-magnitude
relationship; (ii) for young objects, the gravity is small ($\log g \simle 3$) and
sphericity effects in the resolution of the transfer equations might come into play.
The derivation of such complete models is under work, but Figure 2 displays the present
pre-MS models (Baraffe \etal, 1998) based on non-grey plane-parallel atmosphere models in a theoretical HR diagram.
 
\begin{figure}
\begin{center}
\epsfxsize=85mm
\epsfysize=85mm
\epsfbox{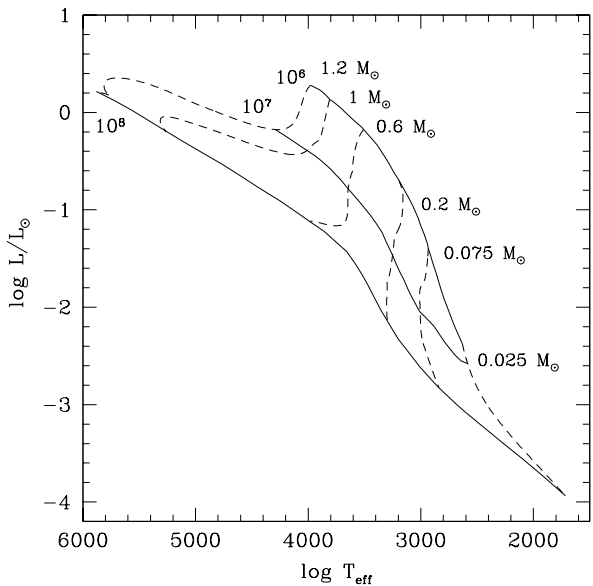}
\end{center}
\caption[]{Pre-main sequence isochrones (models are available from ftp site - see at the end of the paper)}
\label{eps3}
\end{figure}

\section{Stellar and brown dwarf mass functions}

\subsection{Stellar mass function}

The determination of the contribution of LMS and SSOs to the Galactic mass budget requires the correct determination (slope and
normalization) of the mass-function (MF)
down to the hydrogen burning limit, in order to have a solid foundation for extrapolation into
the substellar domain.
This issue, however, is presently not completely settled for the Galactic disk, and significant differences
exist between the MF inferred from parallax-determined luminosity functions (LF) (Kroupa, 1995)
and from the HST
photometric LF (Gould, Flynn \& Bahcall, 1997) (see M\'era, Chabrier \& Schaeffer, 1998,
Fig. 1). It is relatively safe to say, however,
that the MF keeps rising down to the H-burning limit, although with a slope shallower
than a Salpeter MF.
The LMS ($m\le 0.6\,\msol$) mass densities inferred from the integration of the
two afore-mentioned MFs (nearby and HST)
yield, respectively, $\rho_{LMS}\approx (1.9\pm 0.1)\times 10^{-2}$ and $\rho_{LMS}\approx (2.6\pm 0.2)\times 10^{-2}$ $\mvol$ (M\'era \etal, 1998). Adding up the contribution from more massive stars and stellar remnants yields  $\rho_\star \approx (4.0$-$4.6)\pm 0.3
\times 10^{-2}$ $\mvol$ in the Galactic disk, i.e. a surface density $\Sigma_\star\approx 31\pm 2$ $\msurf$. This corresponds to a stellar {\it number}-density $n_\star \sim 0.06$ (HST)
and $\sim 0.35$ pc$^{-3}$ (nearby).

For the spheroid the question is more settled with a MF with $\alpha \simle 1$ (where $dN/dm \propto m^{-\alpha}$) and a stellar
density $\rho_\star < 4.0\times 10^{-5}$ $\mvol$, less than 1\% of the required
dynamical density
(Graff \& Freese, 1996; Chabrier \& M\'era, 1997; Gould \etal, 1998), i.e. a number density
$n_\star < 10^{-3}\, pc^{-3}$.
Extrapolating into the
BD domain yields for the spheroid $n_{BD}< 10^{-4}$ pc$^{-3}$, $\rho_{BD}<\rho_\star/10$, i.e. a microlensing optical depth $\tau\sim 10^{-9}$, about 1\% of the value measured toward the LMC. The puzzle remains unsolved for the {\it dark halo} MF, although both the HDF
observations and the narrow-range of the observed time-distribution of the microlensing
events towards the LMC strongly suggest an IMF different from a Salpeter one below $\sim 1\,\msol$ (see e.g. Chabrier, 1999).

\subsection{Brown dwarf mass function}

A proper census of the number of brown dwarfs ($m<0.07$-$0.09\,\msol$, depending on the metallicity) has significant implications for our understanding of how stars
and planets form.
The determination of the BD MF is a complicated task. By definition BDs never reach
thermal equilibrium and most of the BD's formed at the early stages of the Galaxy will
have fainted to very low-luminosities ($L\propto 1/t$, see e.g. Burrows \& Liebert, 1993).
Thus observations will likely be biased towards young and massive BD's.
The age uncertainty is circumvented when looking for the BD MF in stellar clusters since objects in that case are likely to be coeval. The Pleiades cluster has been extensively surveyed and several BDs have been identified down to $\sim 25\,\mj$ (Martin \etal,
1998; Bouvier \etal, 1998). A single power-law function from $\sim 0.4$ to $0.04\,\msol$ seem to adequately reproduce the observations with some remaining uncertainty in the exponent
$\alpha \sim 0.6-1.0$  (Bouvier \etal, 1998; Martin \etal, 1998). By examinating the various
observations of Doppler radial velocity surveys, Basri and Marcy (1997) reach the conclusion
that all data are consistent with $0\simle \alpha \simle 1$ (see Basri, these proceedings
for an updated analysis of such data). On the other hand Mayor \etal (1997)
find for the mass function of the companions of the G and K dwarf sample observed with CORAVEL
a power-law function with $\alpha \sim 0.4$ from 0.4 down to $0.005\,\msol$. 
The determination of the MF in young clusters is potentially very interesting. Indeed young clusters did not have time to experience evaporation in the outer regions or mass segregation in the
central regions and the present-day MF should reflect relatively closely the initial MF. Such a MF determination, however, is hampered by two observational and theoretical
problems. From the observational point of view, young objects are immersed into dust and a proper
determination of the LF requires a correct determination of the {\it differential} reddening in the cluster. From the theoretical point of view, none of the MFs determined up to now in
young clusters include a careful examination of the effects mentioned in \S3 (initial conditions, sphericity, non-grey effects)
and thus are of dubious validity.
It is thus certainly premature to claim a robust determination of the MF in young clusters.

An independent, powerful information on the stellar and substellar MF comes from microlensing
observations. Indeed,
the time distribution of the events provides a (although model-dependent) determination of
the mass distribution and thus of the minimum mass of the dark objects:
$dN_{ev}/dt_e=E\times \epsilon (t_e)\times d\Gamma/dt_e \propto P(m)/\sqrt m$, where
$E$ is the observed exposure, i.e. the number of star$\times$years, $\epsilon$ is the
experimental efficiency, $\Gamma$ is the event rate and $P(m)$ is the mass probability
distribution.
The analysis of the published 40 MACHO + 9 OGLE events towards the bulge is consistent with a rising mass function at the bottom
of the MS with a minimum mass $m_{inf}\sim 0.05\,\msol$, whereas a decreasing MF
below 0.2 $\msol$ is excluded at the 95\% confidence level (Han \&
Gould, 1995; M\'era \etal, 1998).
Although the time distribution might be affected by various biases
(e.g. blending) and robust conclusions must await for larger statistics,
the present results suggest
that in order to explain
both star counts and the microlensing experiments, a substantial amount of
SSOs must be present in the galactic disk. Indeed, extrapolation of the stellar MF (\S4.1) into the
BD domain down to 0.05 $\msol$ yields for the Galactic disk $\rho_{BD}\sim 4.0\times 10^{-3}\,
\mvol$, i.e. $\Sigma_{BD}\approx 3\,\msurf$, i.e. a BD {\it number}-density comparable to
the stellar one, $n_{BD}\sim 0.1\,pc^{-3}\sim n_\star$.

For the spheroid, extrapolation of the afore-mentioned stellar MF yields $\rho_{BD}\simle  10^{-5}\,\mvol$, $n_{BD}\simle 10^{-4}\,pc^{-3}$, whereas for the dark halo the normalization
is about 2 orders of magnitude smaller (Chabrier \& M\'era, 1997).

\subsection{Planet mass function}

It is obviously very premature to try to infer the mass distribution of exoplanets. This will
first require a clear, both theoretical and observational, distinction between planets and brown dwarfs. An interesting
preliminary result, however, comes from the observed mass distribution of the secondaries
conducted by Mayor and collaborators. As shown by Mayor \etal (1997), there is a
strong discontinuity in the mass distribution at $m_2/sin\, i\approx 5\,\mj$, with a clear
peak below this limit. If confirmed this would suggest that planet formation in a 
protoplanetary disk is a much more efficient mechanism than BD formation, which results
from cloud collapse and fragmentation.

\section{Conclusion and perspectives}

As mentioned in the introduction, accurate models for low-mass stars, brown dwarfs and giant
planets
are needed to shed light on the observable properties of these objects and to
provide guidance to the ongoing and future surveys aimed at revealing their contribution
to the Galactic population. From this point of view, tremendous progress has been made 
in the recent years with the derivation of consistent evolutionary model and synthetic spectra calculations which accurately reproduce
the observed sequences of globular clusters, Pleiades, field objects and of the benchmark BD Gl229B in {\it various photometric passbands} (see references mentioned in the Introduction).
Moreover the afore-mentioned LMS models have been shown to accurately reproduce
the observational mass-magnitude relationship (Henry \& McCarthy, 1993; Henry \etal, 1999) both
in the optical and in the infrared (Baraffe \etal, 1998) and to yield a consistent,
coeval sequence for the quadruple system $GG$-$Tau$, whose component masses extend from
1.2 $\msol$ down to $\sim 0.035\,\msol$ (White \etal, 1999), a formidable test for the theory.
On the other hand, stringent constraints on the theory of dense/cool objects are now provided
by laboratory high-pressure experiments, for the interior, and by various spectroscopic
and photometric observations of LMS and SSO's. Any theory aimed at describing the mechanical and thermal properties of these objects {\it must} be confronted to these experimental/observational constraints in order to assess any degree of validity.
Improvement in the theory will proceed along with the discovery of many more SSO's, hopefully
bridging the gaps on either side of Gl229B from the bottom of the MS to Jupiter-like objects.
Observation of a transit EGP would allow the determination of the radius and the mass of
the object, providing a formidable constraint on the theory. For 51-Peg-like objects, the
probability to observe such a transit $p=d_\star/D$, where $d_\star$ and $D$ are the stellar and orbital diameters, respectively, is $p\sim 10\%$, by no means negligible.

The increasing number of observed LMS and SSOs, together with the derivation of accurate
models, will allow eventually a robust determination of the stellar and substellar mass
functions, and thus of the exact density of these objects in the Galaxy. As discussed in \S4,
present MF determinations in various Galactic regions point to a slowly rising MF near and
below the H-burning limit, with a BD number density comparable to the stellar one. Whether
this behaviour is universal (although we already know it is certainly not the case for the
dark halo), whether it is consistent with a general log-normal form, must await confirmation
from future observations. On the other hand, the amazingly rapid pace of exoplanet discoveries
should yield the determination of the planetary MF. These combined informations will allow
the determination of the BD minimum mass and planet maximum mass.

As seen, the physics of SSOs involves an amazingly large domain of physics and astrophysics,
from the fundamental N-body problem to star formation and Galactic evolution, and
will certainly remain a lively domain of the early 21st century astronomy.

\bigskip
The present grainless models are available from:

\hskip 1cm ftp ftp.ens-lyon.fr, username: anonymous \par
\hskip 1cm ftp $>$ cd /pub/users/CRAL/ibaraffe \par
\hskip 1cm ftp $>$ get BCAH98\_models and BCAH98\_models\_BD \par

\end{document}